\documentclass[procedia]{easychair}

\usepackage{doc}
\usepackage{makeidx}

\newcommand{\Tc}{T_{\mbox{\scriptsize C}}}
\newcommand{\TN}{T_{\mbox{\scriptsize N}}}

\def\procediaConference{99th Conference on Topics of
  Superb Significance (COOL 2014)}

\title{Non-equilibrium behavior of the magnetization\\
 in the helimagnetic phases of the rare earth alloys $R_{1-x}$Y$_{x}$ ($R =$ Gd, Tb, Dy)}

\titlerunning{Non-equilibrium behavior of the magnetization \ldots}

\author{
    Teruo Yamazaki
\and
    Junno Ishiyama
\and
    Yosuke Noya
\and \\
    Mai Kurihara
\and 
    Hiroshi Yaguchi
}

\institute{
  Department of Physics, 
  Faculty of Science and Technology, 
  Tokyo University of Science, 
  Noda, Chiba 278-8510, Japan\\
  \email{t.yamazaki@rs.tus.ac.jp}
 }

\authorrunning{T. Yamazaki, $et. al.$}

\begin{document}

\maketitle


\begin{abstract}


We have performed DC and AC magnetization measurements 
for the rare-earth magnetic alloy systems Gd$_{0.62}$Y$_{0.38}$,  
Tb$_{0.86}$Y$_{0.14}$, and  Dy$_{0.97}$Y$_{0.03}$.
These materials commonly exhibit a proper helical magnetic structure, 
and a ferromagnetic structure at lower temperatures.
In all of these materials, a difference between zero-field-cooled (ZFC) magnetization 
and field-cooled (FC) magnetization and a hysteresis loop in the $M$-$H$ curve 
have been observed in the helimagnetic phases.
The non-equilibrium behavior is possibly caused by a common nature, 
e. g., chiral domain structures.
In addition to the above behavior, 
strong non-linearity of the magnetization and slow spin dynamics 
have been observed around the N\'eel temperature only in Gd$_{0.62}$Y$_{0.38}$.
The spin-glass like behavior observed in Gd$_{0.62}$Y$_{0.38}$ could be 
related to a novel glassy state such as a helical-glass state.
\end{abstract}


%
%

\section{Introduction}
\label{sect:introduction}

The magnetic alloy system Gd$_{1-x}$Y$_{x}$ shows a para-helimagnetic phase transition 
at a N\'eel temperature, $\TN$ 
and a heli-ferromagnetic phase transition at a Curie temperature, $\Tc$ 
in the Y-concentration range of $0.32 < x < 0.40$\cite{Bates}.
In the helimagnetic phase, a proper helical magnetic structure is realized. 
(i.e., The propagation vector is parallel to the $c$-axis and the spin direction is 
perpendicular to the $c$-axis.)
In our previous study, we measured the AC magnetization of Gd$_{0.62}$Y$_{0.38}$, 
and observed the following two kinds of non-equilibrium behavior 
in the helimagnetic phase, 
in spite of the helimagnetic order being 
a long-range antiferromagnetic order \cite{yamazaki}: 
One is an enhancement of the imaginary part of the AC susceptibility 
over the whole helimagnetic temperature range 
and also for temperatures slightly above $\TN$ and below $\Tc$.
The enhancement becomes more remarkable with decreasing measurement frequency between 0.01 and 10 Hz over the helimagnetic temperature range,  
strongly suggestive of anomalous slow dynamics occurring \cite{yamazaki}.
The other is strong non-linearity of the magnetization around $\TN$.
This non-linearity was more clearly observed when the measurement was performed with the magnetic field applied along the $a^{\ast}$-direction 
compare with along the $c^{\ast}$-direction.
On the other hand, such non-equilibrium behavior was not observed in Ho$_{1-x}$Y$_{x}$\cite{yamazaki}.

Tb, Dy, and their diluted alloy systems,Tb$_{1-x}$Y$_{x}$ and Dy$_{1-x}$Y$_{x}$, 
exhibit similar magnetic structures, a proper helimagneic structure for $\Tc < T < \TN$, 
and a ferromagnetic one below $\Tc$, in the low concentration $x$ range \cite{Koehler,Child}.
A rather characteristic hysteresis loop was observed 
in the magnetization curve of Tb$_{0.63}$Y$_{0.37}$ \cite{Nikitin}. 
Recently, temperature hysteresis of the propagation vector has been also observed 
in the helimagnetic phase of Dy\cite{Yu}.
From these experimental facts, intriguing non-equilibrium state is probably realized 
commonly in the helimagnetic phases in these systems.

In this study, we have investigated the DC magnetization of Gd$_{0.62}$Y$_{0.38}$ 
and also the DC and AC magnetization of the analogous materials Tb$_{0.86}$Y$_{0.14}$ and Dy$_{097}$Y$_{0.03}$, 
intending to search for non-equilibrium behavior. 
We have observed a clear difference between field-cooled (FC) and  zero-field-cooled (ZFC) magnetization 
in the helimagnetic phase of Gd$_{0.62}$Y$_{0.38}$, 
and as well as in Tb$_{0.86}$Y$_{0.14}$ and Dy$_{097}$Y$_{0.03}$.
Additionally, we have observed strong non-linearity of the magnetization around $\TN$ 
only in Gd$_{0.62}$Y$_{0.38}$.
We propose that the strong nonlinearity could be 
related to the very weak anisotropy of the Gd$^{3+}$-ion, 
and is possibly due to a glassy state such as a helical-glass state \cite{Ioff, Thomson}.


\section{Experimental}
\label{sect:experimental}

The single-crystalline sample of Gd$_{0.62}$Y$_{0.38}$ used in this study was grown 
by the Czochralski pulling method with a tetra-arc furnace.
The polycrystalline samples of Tb$_{0.86}$Y$_{0.14}$ and Dy$_{0.97}$Y$_{0.03}$ used were prepared 
by arc melting with a mono-arc furnace.
In order to prevent inhomogeneity of the Y-concentration in the samples, quenching was done.
The samples were wrapped in tantalum foils and sealed in evacuated quartz ampoule 
under atmosphere of 25-cmHg argon 
and annealed at 973 K for one week and quenched in iced water.

The DC and AC magnetization was measured using a superconducting quantum interference device (SQUID) magnetometer (MPMS Quantum Design).
The dimensions of Gd$_{0.62}$Y$_{0.38}$ single crystal sample used for measurements 
were 0.68 mm $\times$ 0.15 mm $\times$ 1.34 mm. 
(For magnetization measurements, the applied field was 
along the $a^{\ast}$ direction, corresponding to the direction of the 0.68-mm length.)
Those of Tb$_{0.86}$Y$_{0.14}$ and Dy$_{0.97}$Y$_{0.03}$ polycrystalline samples 
were 4.22 mm $\times$ 0.78 mm $\times$ 0.17 mm 
and 4.55 mm $\times$0.89 mm $\times$ 0.41 mm, respectively. 
(For these samples, the applied magnetic field was 
along the the longest dimension.)
For DC magnetization measurements, we used magnetic field of up to 1 kOe 
since non-equilibrium behavior in Gd$_{0.62}$Y$_{0.38}$ was observed
only at rather low magnetic fields.
For AC magnetization measurements, the applied AC field was 3 Oe 
and the frequency was between 1 and 1000 Hz.
Care was taken to ensure that the remnant field at the sample position was cancelled 
such that the accuracy of the applied field was within 0.2 Oe. 
(We estimated the remnant field by measuring 
the magnetization of a paramagnetic sample of Dy$_{2}$O$_{3}$ at low magnetic fields.)


\section{Results and Discussion}
\label{sect:results}

\begin{figure}[b]
	\begin{centering}
	\includegraphics[width=0.45\textwidth]{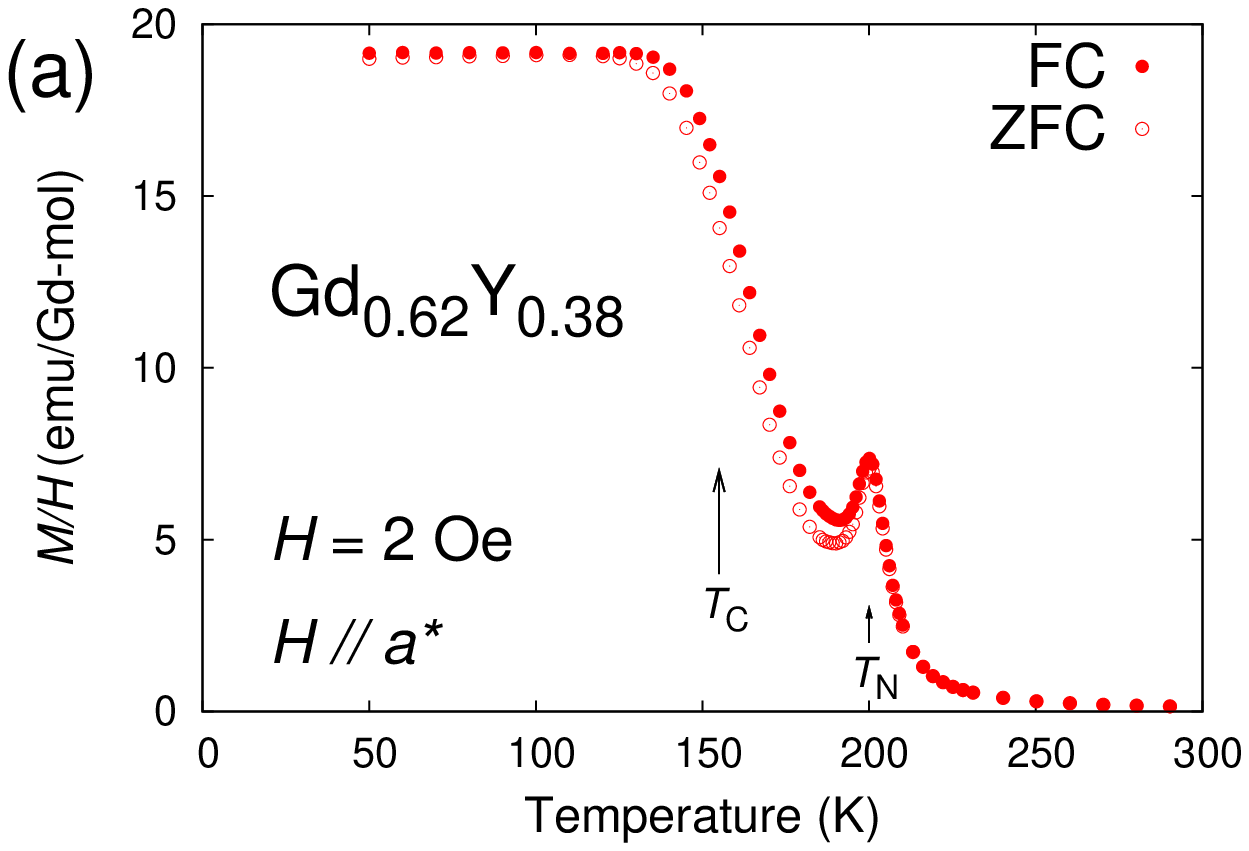}
	\includegraphics[width=0.45\textwidth]{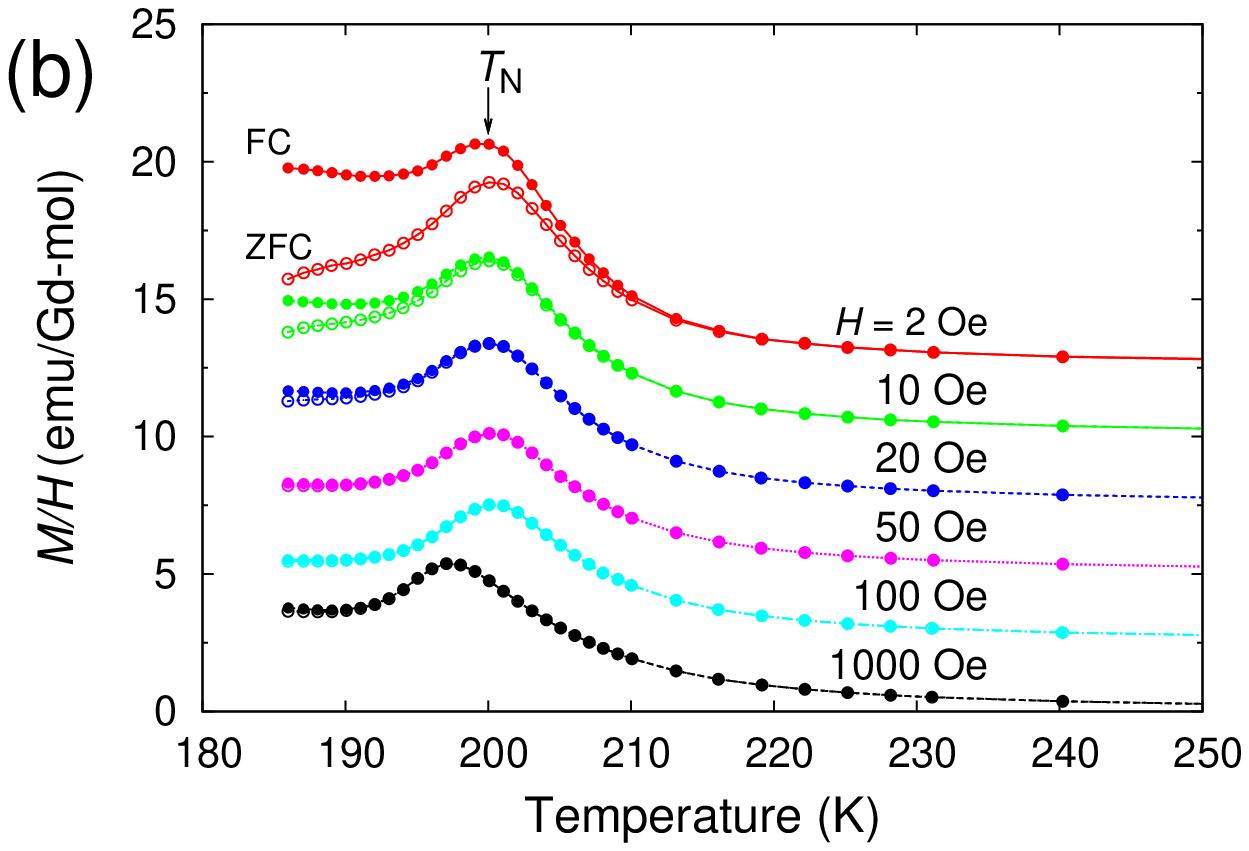}
	\includegraphics[width=0.45\textwidth]{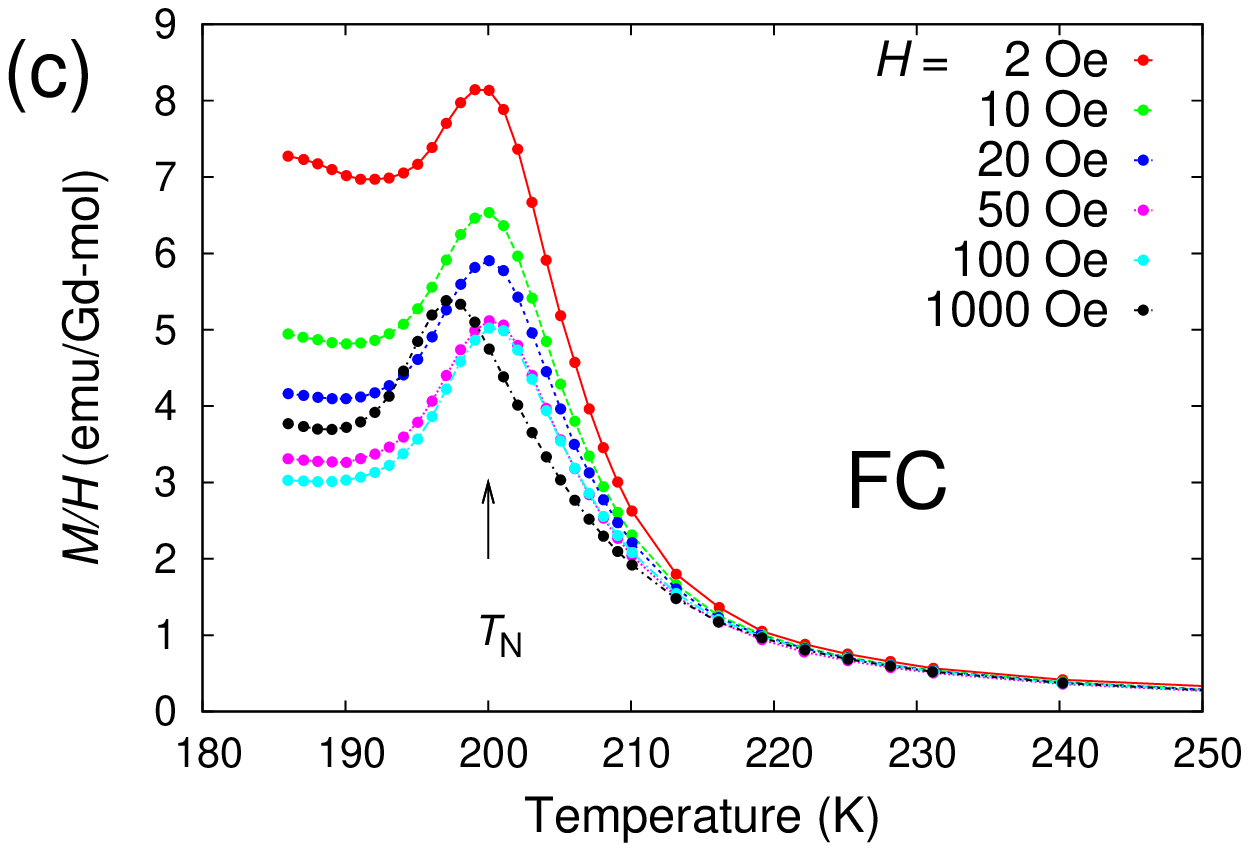}
	\includegraphics[width=0.45\textwidth]{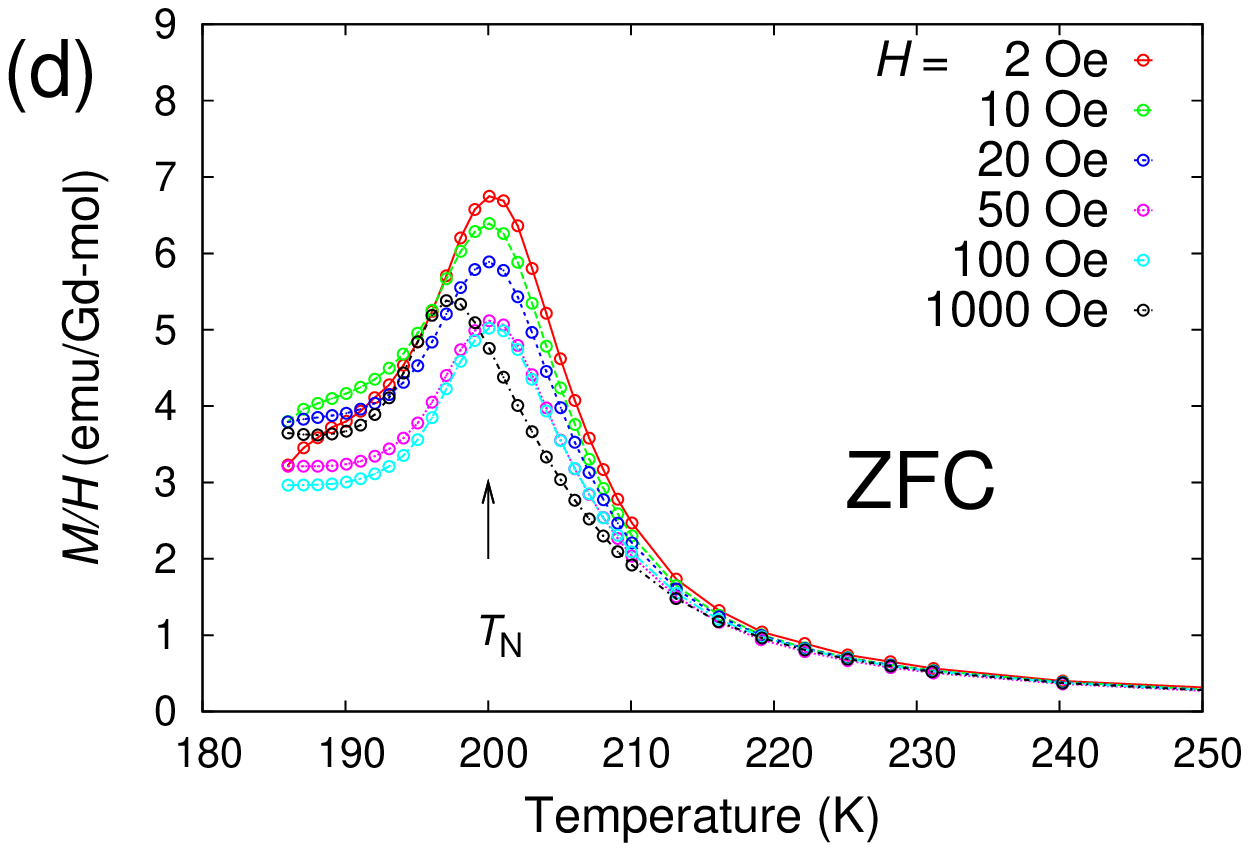}
	\caption{Temperature dependences of the $M/H$ along the $a^{\ast}$ direction of Gd$_{0.62}$Y$_{0.38}$. The measurements were done after cooling the sample down to (a) 50 K and (b)-(d) 186 K  in FC (filled circle) and ZFC (open circle) conditions. 
	In figure (b), traces have been vertically offset by 2.5, 5, 7.5... for visual clarity.
}
	\label{fig:GdY}
	\end{centering}
\end{figure}

    The DC magnetization divided by the magnetic field, $M/H$, along the $a^{\ast}$-direction of Gd$_{0.62}$Y$_{0.38}$ is 
shown as a function of temperature in Figs.\,\ref{fig:GdY} (a)-(d).
The measurements were performed 
in field-cooled (FC) and zero-field-cooled (ZFC) conditions from 300\.K.  
For each condition, data were taken in warming process after cooling the sample 
(a) down to 50~K (well below $\Tc$; i.e, in the ferromagnetic phase), 
and (b) - (d) down to 186 K (between $\Tc$ and $\TN$; i.e., in the helimagnetic phase).
Peaks corresponding to the para-helimagnetic phase transition temperature are clearly seen at $\TN$ of 200 K. 
An increase in magnetization corresponding to the heli-ferromagnetic phase transition 
around $\Tc$ 
is also seen whereas we are unable to determine $\Tc$ using these measurements. 
In Fig.\,\ref{fig:GdY} (a), a marked difference between the FC and ZFC magnetization
was observed 
in the helimagnetic temperature range 
while the FC and ZFC magnetization is almost constant 
in the ferromagnetic temperature range in a low magnetic field of 2 Oe.
The difference between ZFC and FC data was suppressed with increasing field, 
and no difference was observed in fields above 100 Oe.
Moreover, in Fig.\,\ref{fig:GdY} (d), the magnitude of $M/H$ in the ZFC condition is seen 
to be rapidly suppressed by magnetic field of 50 or 100 Oe in particular around $\TN$, 
 indicating strong non-linearity of the magnetization.  
 In higher magnetic fields, the peak associated with the para-helimagnetic phase transition 
at $\TN$ is shifted to lower temperatures 
and $M/H$ at the peak is considerably enhanced at 1000 Oe again.

\begin{figure}[b]
	\begin{centering}
	\includegraphics[width=0.45\textwidth]{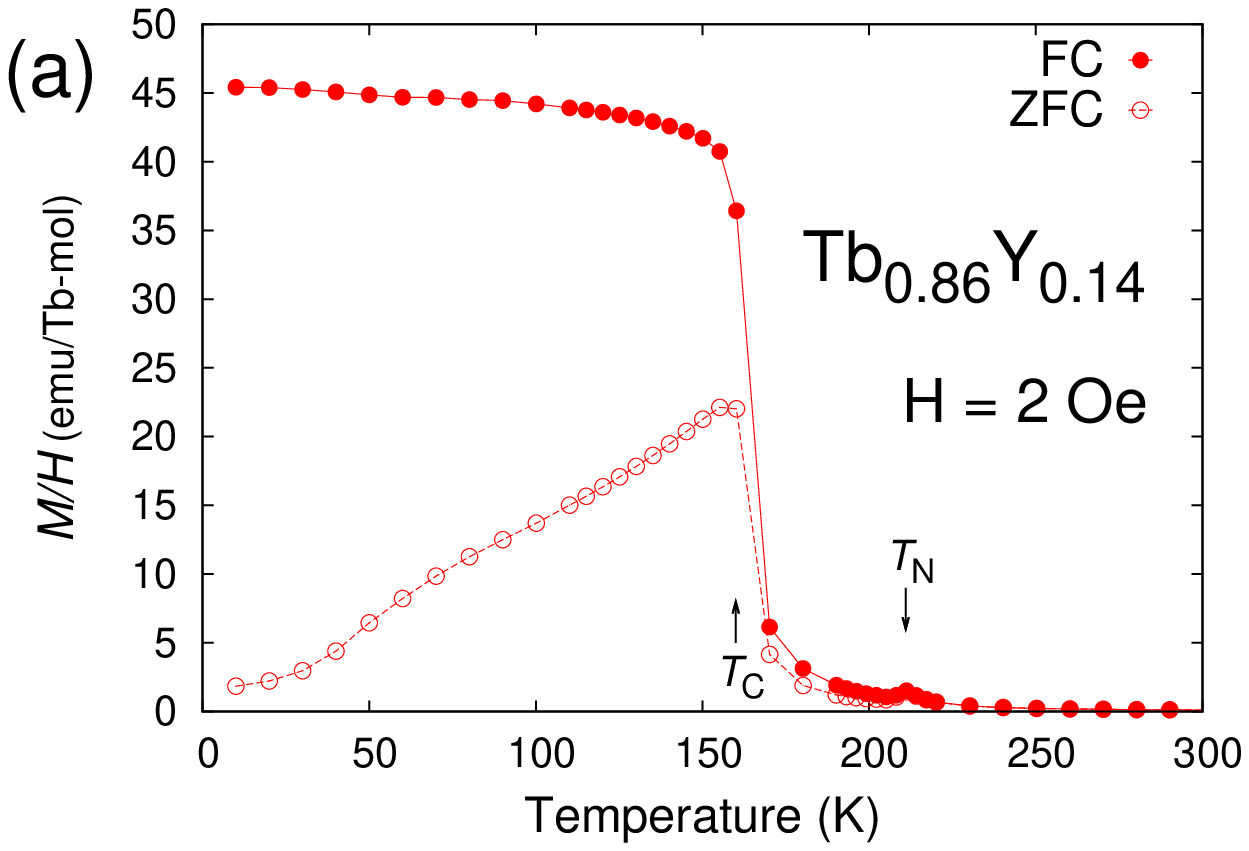}
	\includegraphics[width=0.45\textwidth]{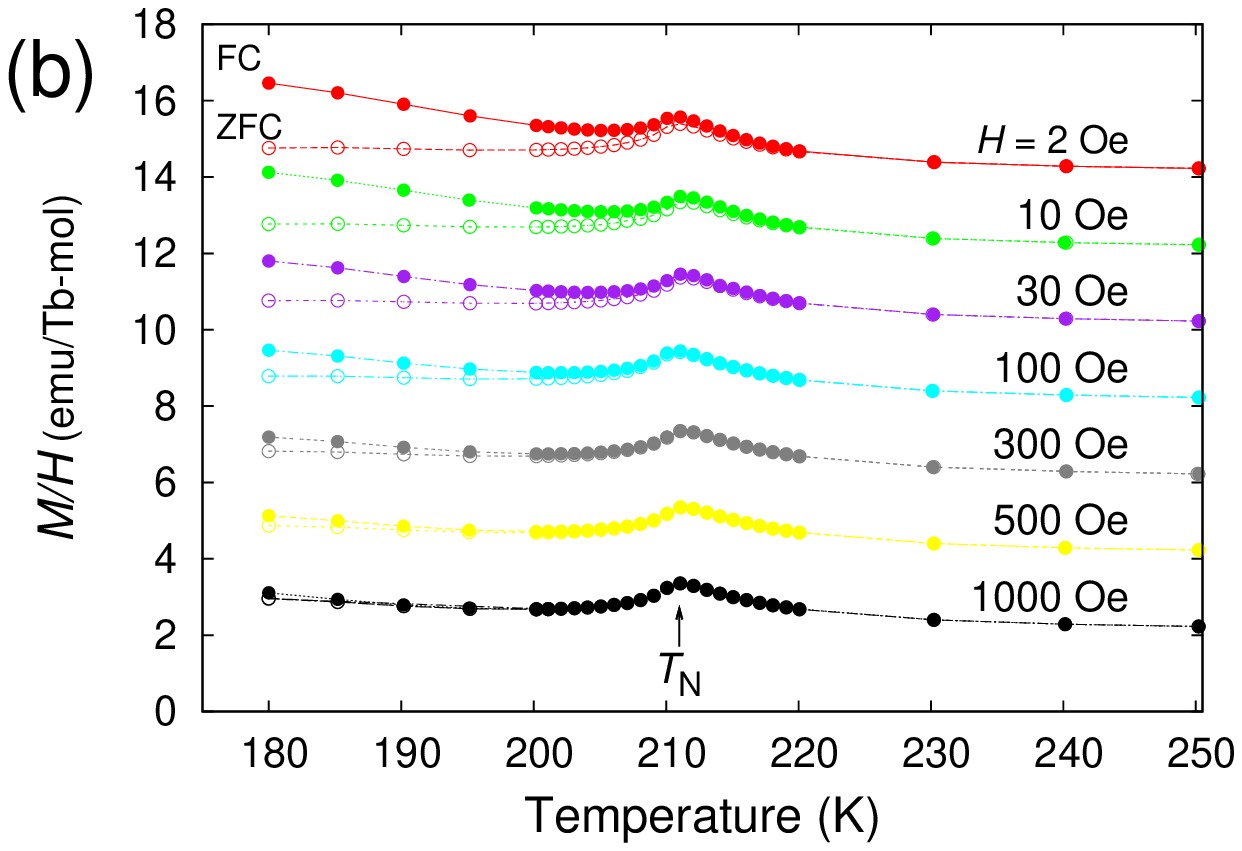}
	\includegraphics[width=0.45\textwidth]{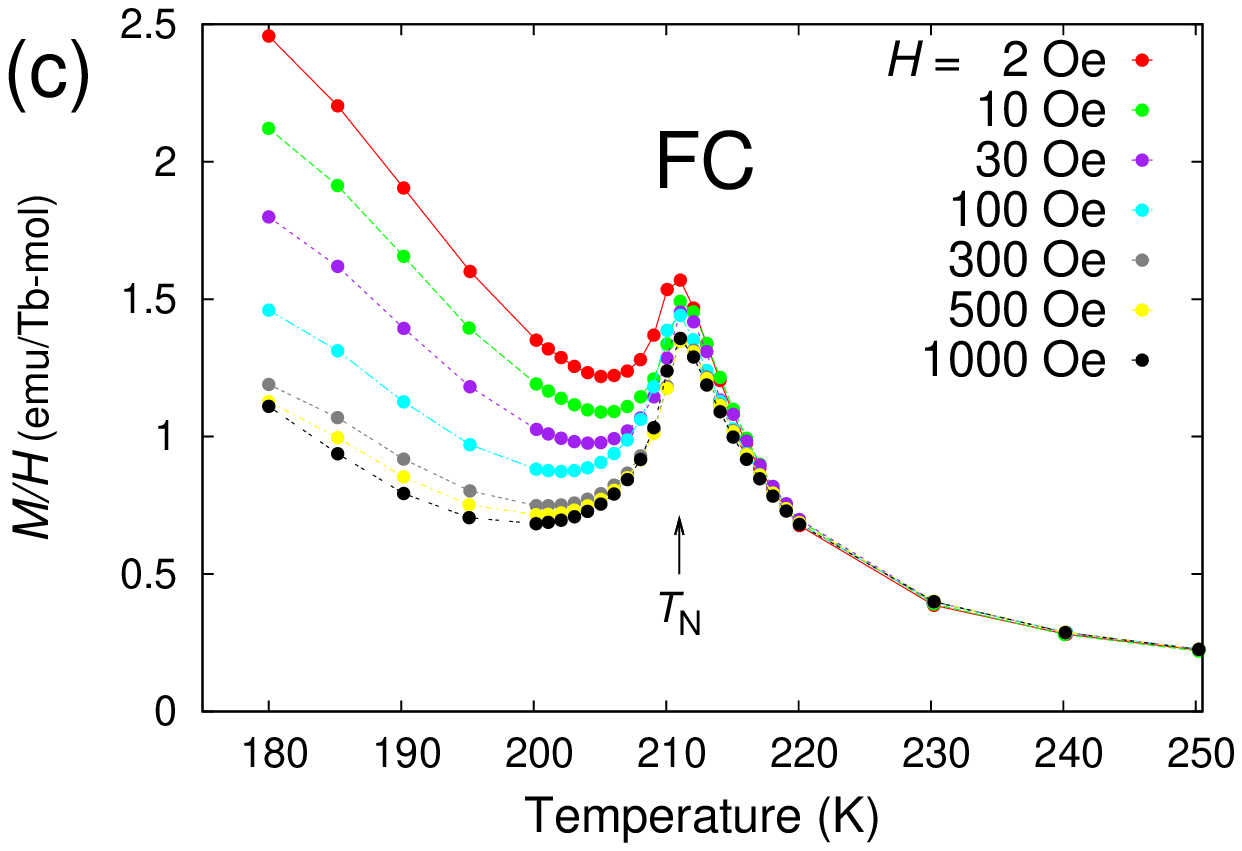}
	\includegraphics[width=0.45\textwidth]{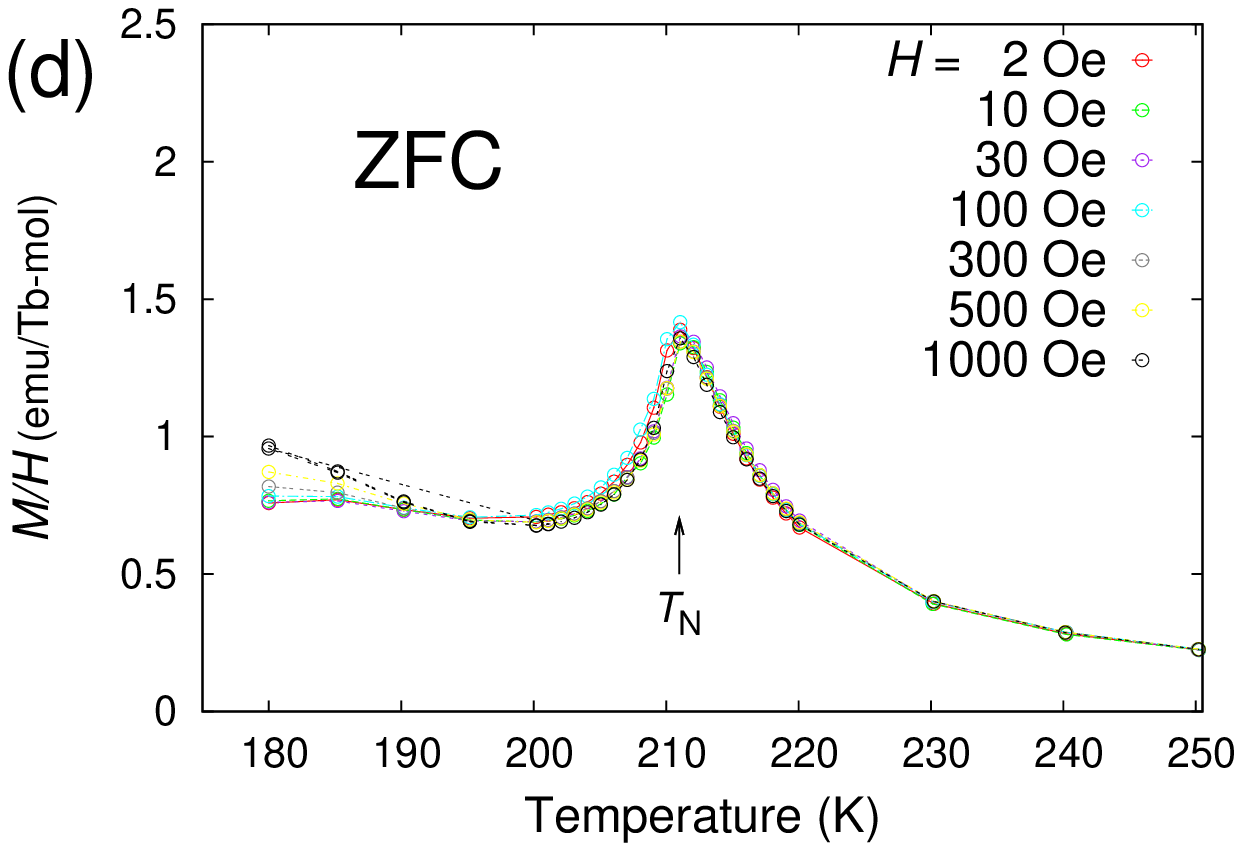}
	\includegraphics[width=0.45\textwidth]{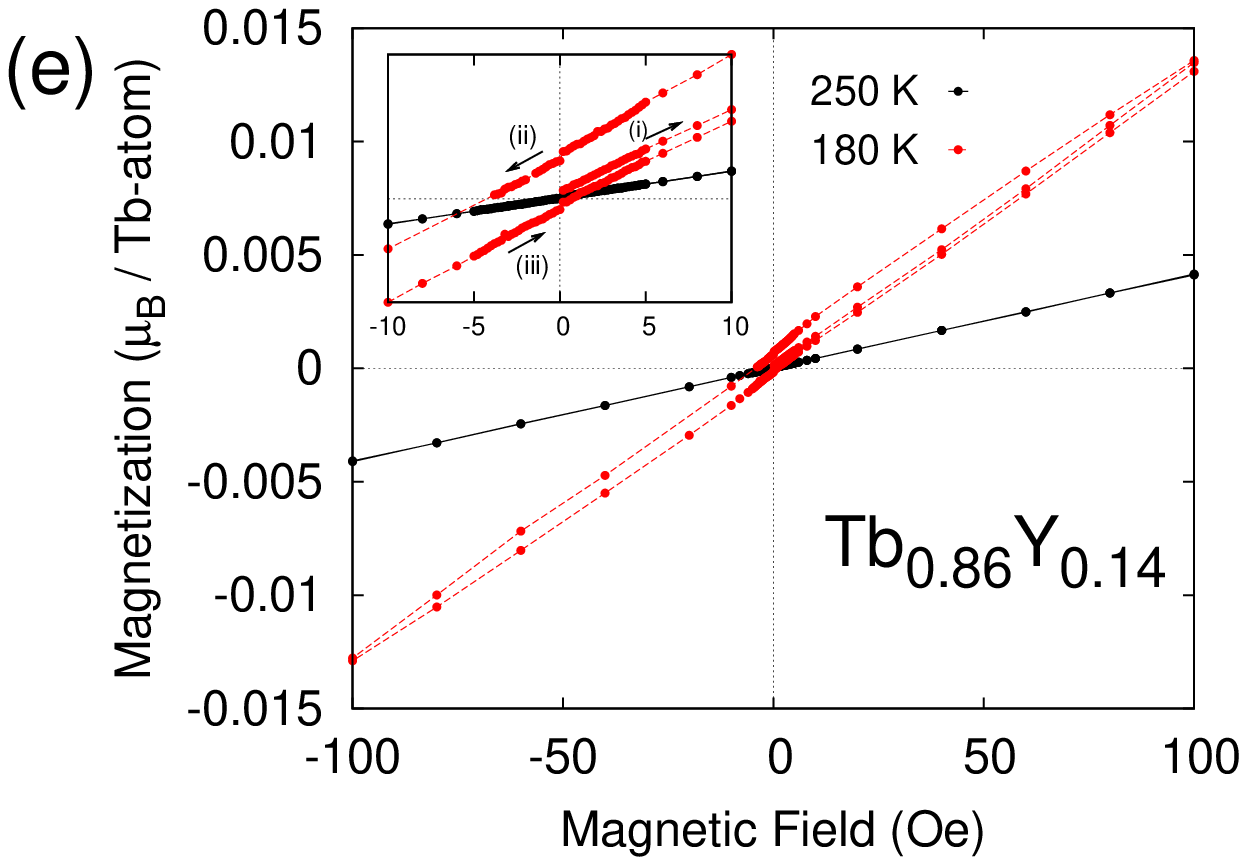}
	\caption{(a)-(d) $M/H$ of Tb$_{0.86}$Y$_{0.14}$ plotted against temperature. The $M/H$-$T$ measurements were done after cooling the sample down to (a) 10 K and (b)-(d)180 K  in  FC (filled circle) and ZFC (open circle) conditions. 
	In figure (b), traces have been vertically offset by 2, 4, 6... for visual clarity.
	(e) magnetization curve of Tb$_{0.86}$Y$_{0.14}$. The $M$-$H$ measurements performed after cooling the sample from 300 K to 180 K or 250 K in zero field.}
	\label{fig:TbY-DC}
	\end{centering}
\end{figure}

\begin{figure}[b]
	\begin{centering}
	\includegraphics[width=0.45\textwidth]{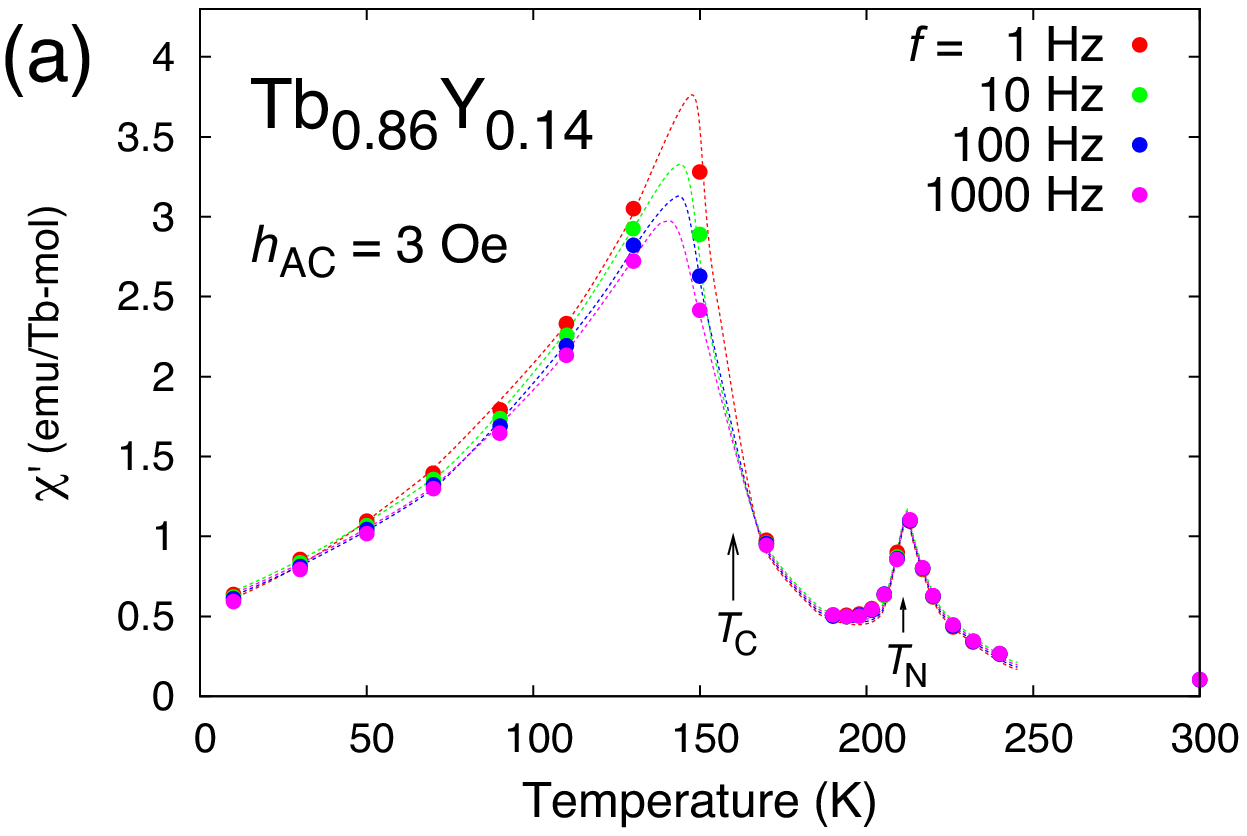}
	\includegraphics[width=0.45\textwidth]{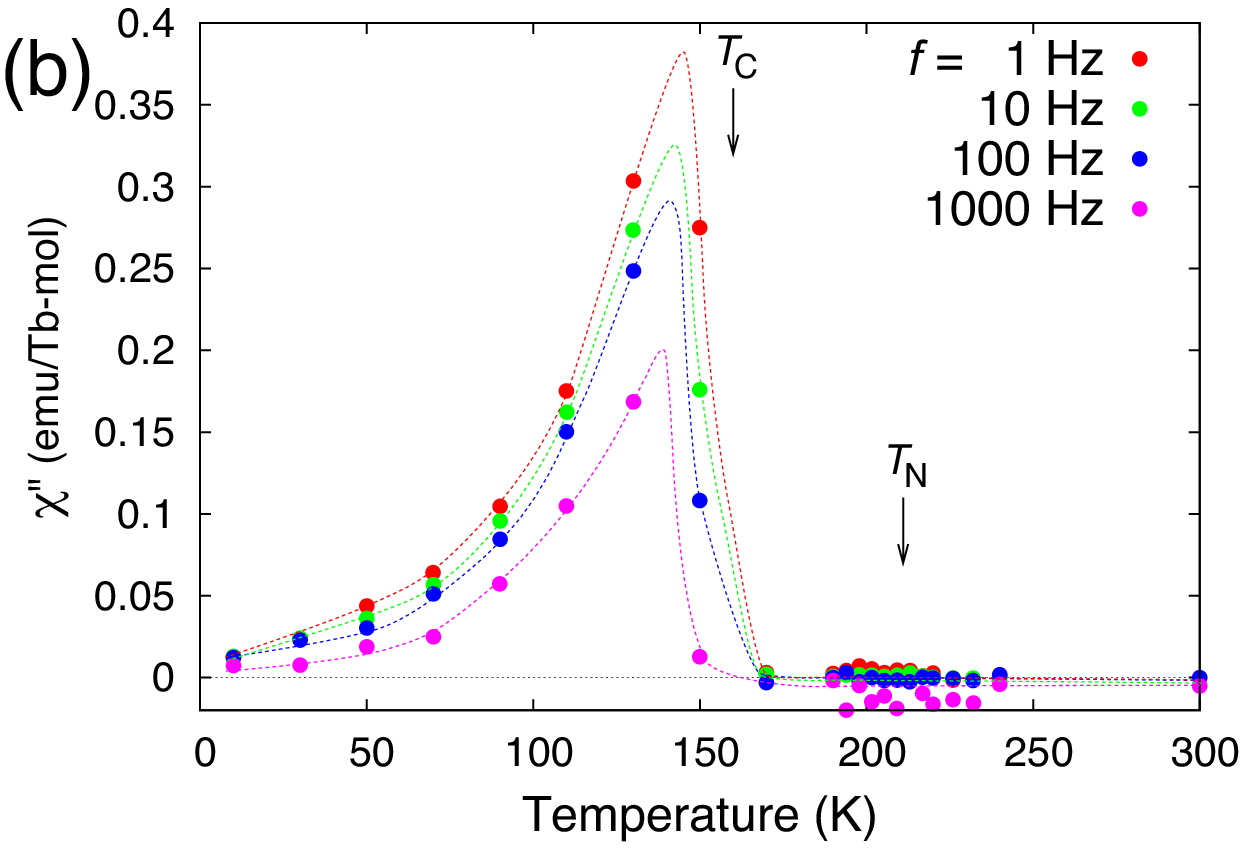}
	\caption{Temperature dependences of  the (a) real and (b) imaginary parts of the AC susceptibility, $\chi'$ and $\chi''$, of Tb$_{0.86}$Y$_{0.14}$. 
	The dotted curves through data points provide guides to the eye.}
	\label{fig:TbY-AC}
	\end{centering}
\end{figure}

\begin{figure}[b]
	\begin{centering}
	\includegraphics[width=0.45\textwidth]{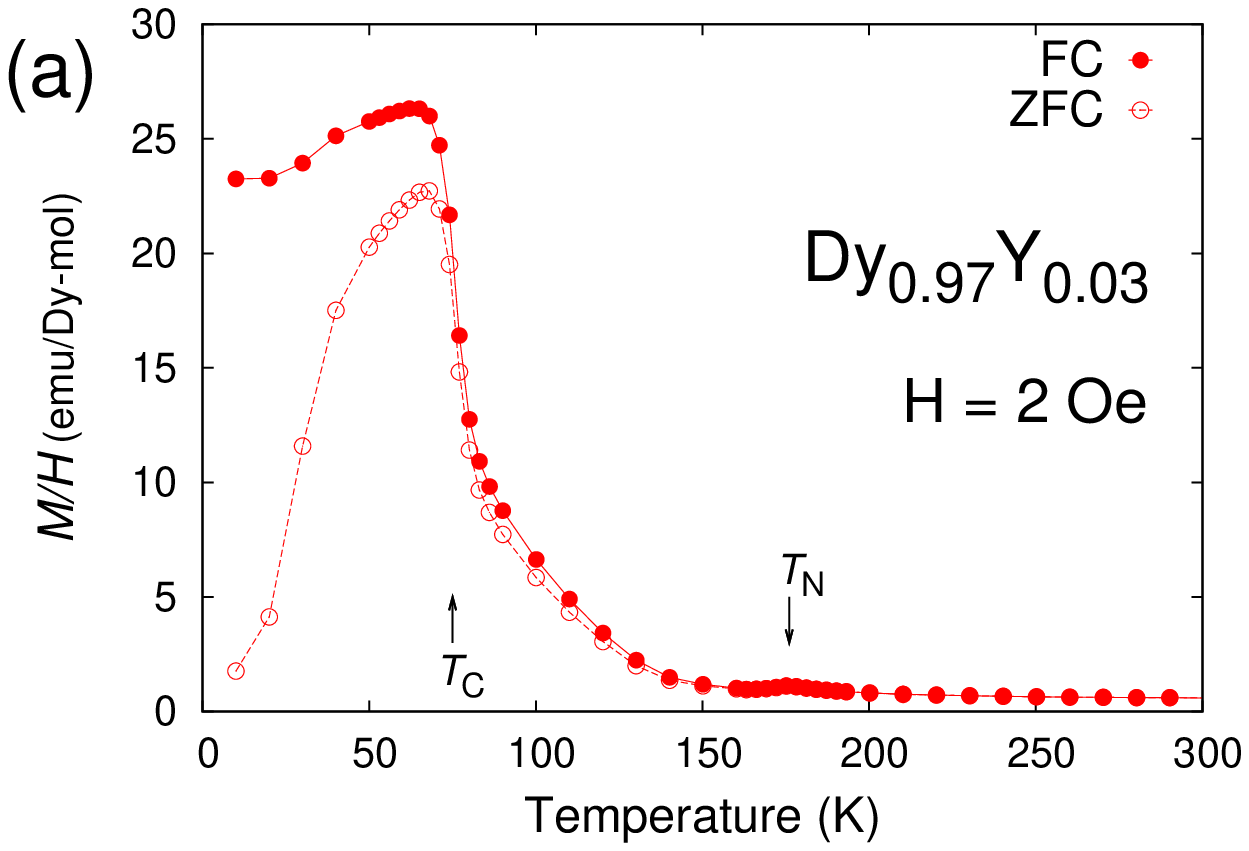}
	\includegraphics[width=0.45\textwidth]{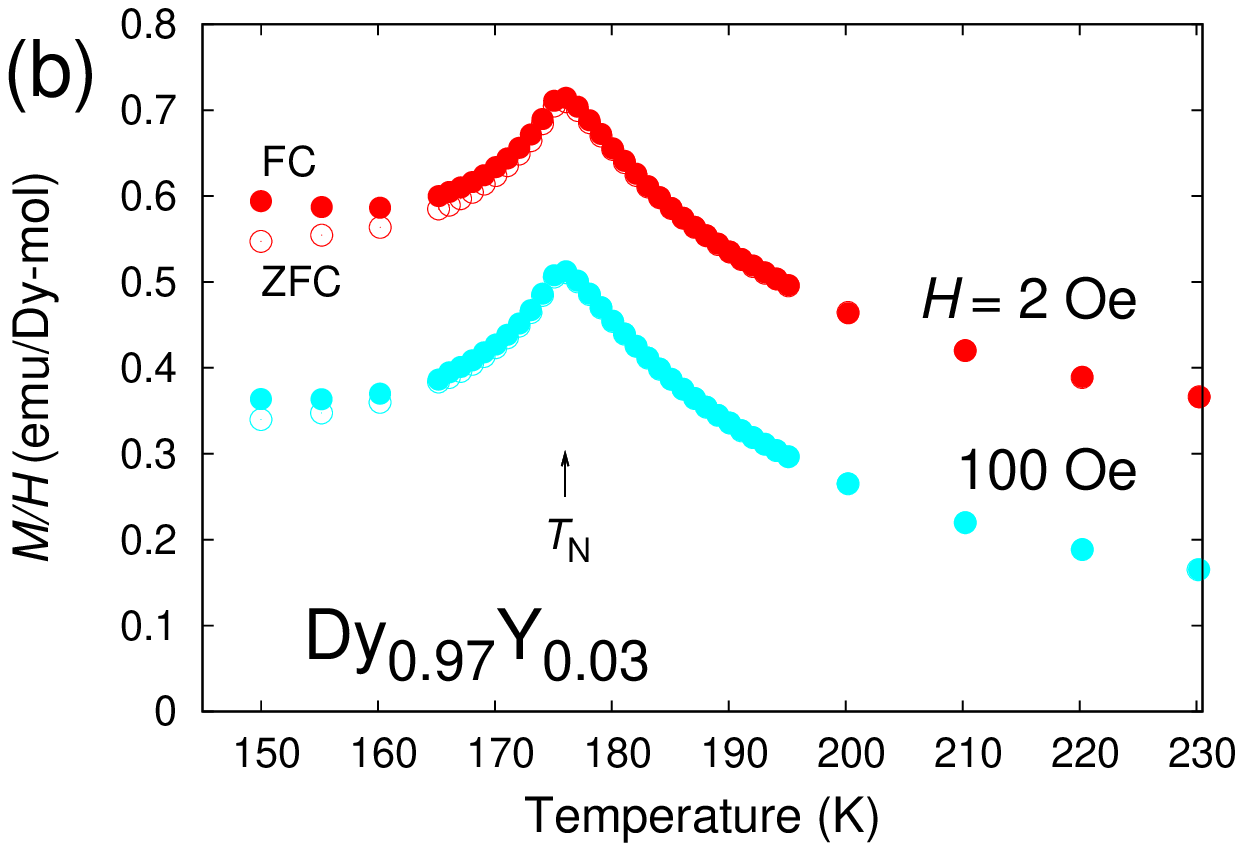}
	\includegraphics[width=0.45\textwidth]{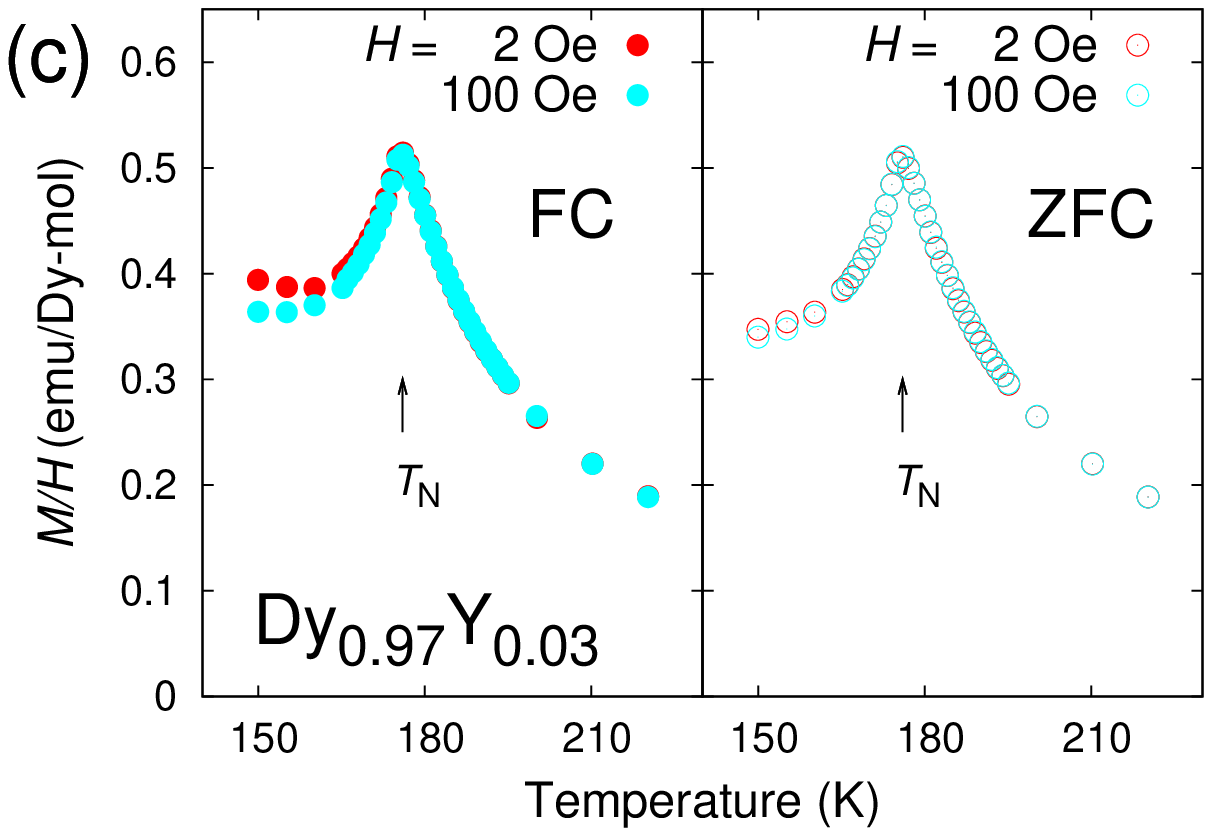}
	\includegraphics[width=0.45\textwidth]{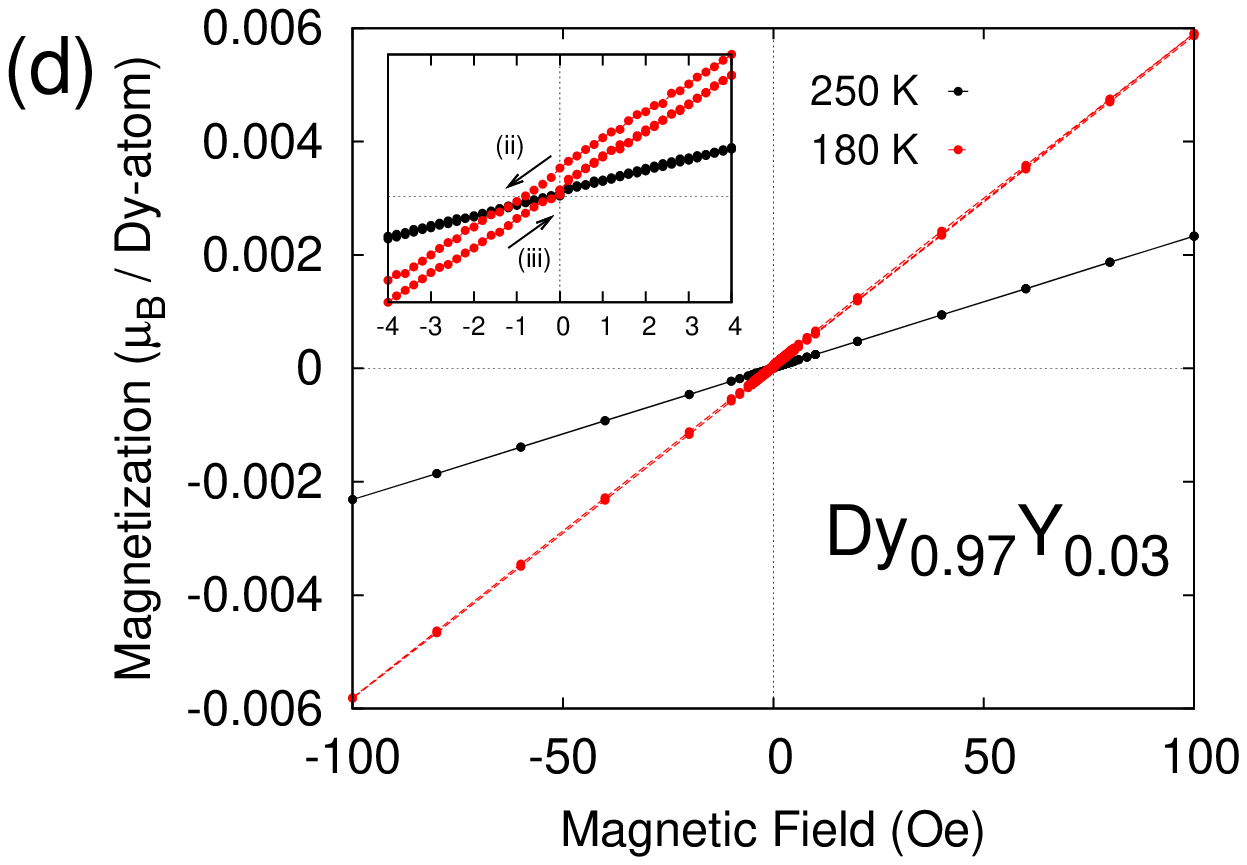}
	\caption{Temperature dependences of $M/H$ of Dy$_{0.97}$Y$_{0.03}$ measured after cooling the sample (a) down to 10 K, and (b) and (c) down to 150 K  in  FC (filled circle) and ZFC (open circle) conditions. In figure (b), traces have been vertically offset by 0.2 for visual clarity.
(d) magnetization curve measured for Dy$_{0.97}$Y$_{0.03}$ after cooling the sample from 300 K to 150 K or 250 K in zero field.
}
	\label{fig:DyY}
	\end{centering}
\end{figure}

We have also performed similar measurements on Tb$_{0.86}$Y$_{0.14}$
(Figs.\,\ref{fig:TbY-DC} (a), (c), and (d)).
The measurements were performed in FC and ZFC conditions from 300K. 
Data were acquired in warming process after cooling the sample 
(a) down to 10 K (well below $\Tc$; i.e, in the ferromagnetic phase), 
and (b)-(d) down to 186 K (between $\Tc$ and $\TN$; i.e., in the helimagnetic phase).
In Fig.\,\ref{fig:TbY-DC} (a), 
peaks corresponding to the para-helimagnetic phase transition at $\TN$ of 211 K 
are seen in similar manners for both of the FC and ZFC measurements. 
By contrast, the way an increase in magnetization corresponding to heli-ferromagnetic phase transition 
occurs around $\Tc$ differs between the FC and ZFC data. 
In the ferromagnetic temperature range, the ZFC magnetizaion 
is significantly smaller than the FC magnetization.
With decreasing temperature, the ZFC magnetization suddenly increases around $\Tc$, 
and gradually decreases.
This is typical behavior of ferromagnets involving domain formation process in low magnetic field.
Moreover, in Fig.\,\ref{fig:TbY-DC} (b), 
a large difference between the FC and ZFC magnetization is also evident  
in the helimagnetic temperature range, 
whereas it is smaller than that in the ferromagnetic temperature range 
as seen in Fig.\,\ref{fig:TbY-DC} (a).

$M/H$ measured in the FC condition is strongly suppressed with increasing magnetic field below $\TN$.
On the other hand, $M/H$ in the ZFC condition hardly changes with magnetic field, 
and the linearity of the magnetization is preserved around $\TN$ (see Fig.\,\ref{fig:TbY-DC} (d)).
The magnetization curves obtained at 180 and  250 K are shown in Fig.\,2 (e).
The measurements were performed by changing the magnetic field 
in the sequence of (i) 0\,Oe $\rightarrow$ 100\,Oe, 
(ii) 100\,Oe $\rightarrow$ $-100$\,Oe,  and (iii) $-100$\,Oe   $\rightarrow$ 100\,Oe.
At 180 K, in the helimagnetic phase, a hysteresis loop with a finite coercivity was observed.
This hysteresis loop should be closely related to 
the suppression of $M/H$ in the FC condition by magnetic field below $\TN$ 
seen in Fig.\,\ref{fig:TbY-DC} (c). 
The ZFC condition in the $M/H$-$T$ measurements corresponds to the process 
(i) in the $M$-$H$ curve.
In this process, the linearity is preserved and $M/H$ in the ZFC condition remains constant.
On the other hand, the state in the FC condition in the $M/H$-$T$ measurements 
should be close to that of the process (ii) in the $M$-$H$ measurements.
In this process, $M/H$
 should become larger as the applied field approaches to zero 
owing to the finite remnant magnetization.
In other words, the difference between the ZFC and FC magnetization is observed because of the finite coercivity.

The real and imaginary parts of the AC susceptibility, $\chi'$ and $\chi''$,  of Tb$_{0.86}$Y$_{0.14}$ 
are plotted against temperature in Fig.\,\ref{fig:TbY-AC} (a) and (b), respectively.
Peaks corresponding to $\Tc$ and $\TN$ were observed in the real part of the AC susceptibility. 
In the imaginary part of the susceptibility, 
peaks around $\Tc$ were observed, 
suggestive of critical slowing down in common to Gd$_{0.62}$Y$_{0.38}$ \cite{yamazaki}.
However, no discernible enhancement of the imaginary part $\chi''$ around $\TN$ 
was observed in Tb$_{0.86}$Y$_{0.14}$. 
This contrasts with Gd$_{0.62}$Y$_{0.38}$, 
in which an enhancement of the imaginary part $\chi''$ was observed around $\TN$
as mentioned in Introduction \cite{yamazaki}.

The temperature dependence of $M/H$ of Dy$_{0.97}$Y$_{0.03}$ 
is shown in Figs.\,\ref{fig:DyY}\,(a)-(c).
The measurements were performed in warming process after cooling the sample down to the ferromagnetic temperature of 10\,K from 300\,K (Fig.\,\ref{fig:DyY}(a)), and down to the helimagnetic temperature of 150\,K in FC and ZFC conditions from 300\,K (Figs.\,\ref{fig:DyY}(b),(c)).
Peaks corresponding to the para-helimagnetic phase transition at {$\TN$} of 176\,K and an increase of magnetization corresponding to heli-ferromagnetic phase transition at {$\Tc\sim$} 75\,K are seen.
In Fig.\,\ref{fig:DyY}~(b), a difference between the FC and ZFC magnetization below $\TN$
was also observed, 
although it was smaller than that observed in Tb$_{0.86}$Y$_{0.14}$.
A suppression of $M/H$ in the FC condition with increasing magnetic field was also observed, 
and that in the ZFC condition was hardly observed in Fig.\,\ref{fig:DyY}\,(c).
The magnetization curves measured after cooling the sample from 300\,K  to 150\,K or 250\,K in zero field are shown in Fig.\,\ref{fig:DyY}~(d).
A hysteresis loop with a finite coercivity was also observed at the helimagnetic phase temperature of 150\,K. 
The manner of the non-equilibrium behavior was qualitatively the same 
as that in Tb$_{0.86}$Y$_{0.14}$.
Therefore, the difference between the FC and ZFC magnetization 
and the suppression of $M/H$ in the FC condition by magnetic field observed in Dy$_{0.97}$Y$_{0.03}$ 
is probably owing to the appearance of a finite coercivity in the helimagnteic phase 
as well as in Tb$_{0.86}$Y$_{0.14}$.

Let us finally discuss possible origins of the non-equilibrium behavior 
observed in the helimagnetic phase of Gd$_{0.62}$Y$_{0.38}$,  Tb$_{0.86}$Y$_{0.14}$, 
and Dy$_{0.97}$Y$_{0.03}$, 
placing emphasis on behavior of Gd$_{0.62}$Y$_{0.38}$ among them.

The non-equilibrium behavior commonly observed in the helimagnetic phases 
of all of these materials may be itemized as follows:
(1) a difference between the FC and ZFC magnetization, 
 and (2) a suppression of $M/H$ in the FC condition with increasing applied magnetic field.
As stated above, the behavior should be 
related to the appearance of the finite coercivity 
and the hysteresis loop seen in Figs.\,\ref{fig:TbY-DC}\,(e) and \ref{fig:DyY}\,(d).
Non-equilibrium behavior in physical properties in helimagnetic phases in rare-earth alloys 
has been, from quite a while before, discussed in terms of a helical domain structure,
which is formed by the left- and right-handed spin chirality domains, 
in ref. \cite{Palmer, Baruchel}.   
In ref. \cite{Palmer}, hysteretic behavior of the elastic constant $C_{33}$ of Tb$_{0.50}$Ho$_{0.50}$ was observed in the up-sweep-field process and down-sweep-field process,
 and the value of $C_{33}$ at zero field was changed from the initial one.
Consequently, it is naturally inferred that 
the hysteresis loop and the finite coercivity in the $M$-$H$ curve 
could be caused by an influence of the chiral domain structure as well.
Alternatively, the propagation wave vector \mbox{\boldmath $k$} 
(i. e., the period of the helix) could be changed after the application of magnetic field. 
In fact, recent neutron scattering experiments for Dy have observed 
temperature hysteresis of \mbox{\boldmath $k$} \cite{Yu}.

In addition to the above mentioned non-equilibrium behavior (1) and (2), 
we have observed the following non-equilibrium behavior 
only in the Gd$_{0.62}$Y$_{0.38}$ in the present study.
(3) strong non-linearity of the magnetization around $\TN$, 
(4) an enhancement of the imaginary part of the AC susceptibility 
over the whole helimagnetic phase temperature range \cite{yamazaki}, 
indicating anomalous slow dynamics.
These features are similar to those observed generally in spin-glass systems.
This similarity suggests that a novel state such as a helical-glass state
that is theoretically proposed might be realized 
in the helimagnetic phase of Gd$_{0.62}$Y$_{0.38}$\cite{Ioff, Thomson}.
We propose that the spin-glass-like behavior in the Gd$_{1-x}$Y$_{x}$ 
could be closely related to 
the very weak magnetic anisotropy of Gd$^{3+}$-ion 
owing to lack of the orbital angular momentum, $L=0$.

In summary, 
we have investigated the DC and AC magnetizations of the rare-earth alloy systems 
Gd$_{0.62}$Y$_{0.38}$,  Tb$_{0.86}$Y$_{0.14}$,  and Dy$_{0.97}$Y$_{0.03}$. 
All of these materials commonly have a helimagnetic phase 
and, at lower temperatures, a ferromagnetic phase.
We have observed several kinds of non-equilibrium behavior in the helimagnetic phase.
A difference between the FC and ZFC magnetization 
and a hysteresis loop with a finite coercivity,  which are rather unusual, 
should be caused by the same nature, e. g., helical domain structures, 
in the systems.
Among them, strong non-linearity of the magnetization around $\TN$ 
and slow dynamics were particularly unusual, 
and observed only in Gd$_{0.62}$Y$_{0.38}$. 
These features are somewhat reminiscent of a spin-glass, 
and perhaps could be accounted for 
by the formation of a helical glass state 
due to the very weak anisotropy of the Gd$^{3+}$-ion.

\subsection*{Acknowledgement}
We thank Y. Tabata and H. Nakamura for helpful discussions, 
and K. Motoya for allowing us to use equipment for sample synthesis.



%

\end{document}